\documentclass[a4paper,prx,reprint,twocolumn,superscriptaddress]{revtex4-1} 

\usepackage{amsmath,amssymb,amsfonts}
\usepackage{bbold}
\usepackage{graphicx}
\usepackage{color,xcolor}
\usepackage{mathrsfs}
\usepackage{tabularx}
\usepackage{pifont}
\usepackage{verbatim}
\usepackage{psfrag}
\usepackage{mathcomp}
\usepackage{ulem}
\usepackage[colorlinks,citecolor=blue,urlcolor=blue]{hyperref}

\begin{document}

\title[Article Title]{Direct Nuclear-Level Qubits using Trapped \({}^{229}\mathrm{Th}^{3+}\) Ions: A Platform for Entanglement and Universal Quantum Information Processing}

\author{Jingbo Wang}
\affiliation{Beijing Academy of Quantum Information Sciences, Beijing 100193, China}
\affiliation{State Key Laboratory of Low Dimensional Quantum Physics, Department of Physics, Tsinghua University, Beijing, 100084, China}

\author{Haixing Miao}
\affiliation{State Key Laboratory of Low Dimensional Quantum Physics, Department of Physics, Tsinghua University, Beijing, 100084, China}
\affiliation{Frontier Science Center for Quantum Information, Beijing 100184, China}

\author{Shiqian Ding}
\email{dingshq@mail.tsinghua.edu.cn}
\affiliation{State Key Laboratory of Low Dimensional Quantum Physics, Department of Physics, Tsinghua University, Beijing, 100084, China}
\affiliation{Beijing Academy of Quantum Information Sciences, Beijing 100193, China}
\affiliation{Frontier Science Center for Quantum Information, Beijing 100184, China}

\author{Dong E. Liu}
\email{dongeliu@mail.tsinghua.edu.cn}
\affiliation{State Key Laboratory of Low Dimensional Quantum Physics, Department of Physics, Tsinghua University, Beijing, 100084, China}
\affiliation{Beijing Academy of Quantum Information Sciences, Beijing 100193, China}
\affiliation{Frontier Science Center for Quantum Information, Beijing 100184, China}
\affiliation{Hefei National Laboratory, Hefei 230088, China}

\begin{abstract}
The low-energy isomeric transition in Thorium-229 offers a unique interface between nuclear and atomic physics, presenting a resource for quantum technologies that is notably resilient to environmental decoherence. While early experiments focused on nuclei in solid-state crystals~\cite{tiedau2024laser,hiraki2024controlling,zhang2024frequency}, the recent advent of a continuous-wave vacuum ultraviolet laser at 148.4~nm~\cite{xiao2025cwVUV} now enables direct coherent control of individual trapped \({}^{229}\mathrm{Th}\) ions. Building on this breakthrough, we present a theoretical framework for utilizing trapped \({}^{229}\mathrm{Th}^{3+}\) ions as high-fidelity nuclear-level qubits, wherein quantum state preparation, single-qubit control, and entangling operations based on nuclear energy levels can all be efficiently realized. We analyze a scheme to generate entanglement between the nuclear isomeric states of two ions through phonon-mediated coupling, driven by optimized red- and blue-detuned laser sideband pulses. Our analysis, grounded in realistic experimental parameters, also demonstrates that high-fidelity entanglement is achievable, leveraging the nucleus's intrinsically long coherence times. These results provide a practical roadmap for developing nuclear-based quantum information processors and suggest that entangled nuclear-level qubits could potentially unlock new frontiers in precision metrology.
\end{abstract}



\maketitle

\section{Introduction}\label{sec1}
The development of optical atomic clocks has achieved fractional uncertainties below $10^{-18}$~\cite{campbell2012single, tong2025ticking}, yet environmental perturbations and atomic systematics impose fundamental limits. The $^{229}$Th nucleus offers a revolutionary alternative through its unique low-lying isomeric state at $\sim$8.4~eV---uniquely accessible to laser techniques among nuclear transitions~\cite{flambaum2006enhanced,beck2007energy,beeks2021thorium,seiferle2022extending}. First predicted in the 1970s~\cite{kroger1976features}, this transition's extreme narrow linewidth and insensitivity to electromagnetic perturbations~\cite{peik2003nuclear} position it as an ideal candidate for next-generation frequency standards, quantum information processing, and fundamental physics tests~\cite{rellergert2010constraining,fadeev2020sensitivity,peik2021nuclear}.

Decades of progress have yielded increasingly precise measurements of the $^{229}$Th nuclear transition properties and energy~\cite{helmer1994excited,guimaraes2005energy,thielking2018laser}. Early works relied on indirect nuclear physics experiments~\cite{seiferle2019energy,masuda2019x,yamaguchi2019energy,sikorsky2020measurement} until the first direct observation of the isomeric decay~\cite{von2016direct}. This breakthrough triggered rapid advances in theoretical and experimental control, leading to more accurate energy measurements~\cite{scharl2023setup,elwell2024laser}. A key development was the doping of $^{229}$Th into transparent crystals, creating solid-state ensembles with large numbers of nuclei~\cite{tiedau2024laser,hiraki2024controlling,zhang2024frequency}. These high-density platforms enable rapid signal acquisition and sophisticated laser interrogation schemes, including four-wave mixing (FWM)~\cite{tiedau2024laser, elwell2024laser} and vacuum ultraviolet (VUV) frequency combs~\cite{diddams2020optical, zhang2024frequency}. Most recently, these ensembles have allowed direct laser excitation of the isomer and precise frequency ratio measurements against optical atomic clocks, significantly improving measurement precision.

The extreme rarity of \({}^{229}\mathrm{Th}\) makes acquiring the large number of atoms needed for solid-state ensembles particularly expensive. The required nuclei are synthesized from the \(\alpha\)-decay of \({}^{233}\mathrm{U}\)~\cite{kroger1976features} or \(\beta\)-decay of isotopes like \({}^{229}\mathrm{Ac}\)~\cite{von2016direct} and \({}^{229}\mathrm{Pa}\)~\cite{griswold2018production}, then implanted into host crystals such as CaF$_2$~\cite{hogle2016reactor}. Beyond this production challenge, the solid-state approach is intrinsically flawed. The nuclear transition’s sensitivity to environmental variations in charge density and field gradients~\cite{kazakov2012performance,peik2015nuclear,higgins2025temperature} causes inhomogeneous broadening. This, combined with uncontrolled local fields and the inability to address single nuclei, prevents quantum control and multi-nuclear entanglement, thus hindering the development of nucleus clocks and scalable quantum information protocols.

To overcome these challenges, an alternative strategy is to trap individual ionized \({}^{229}\mathrm{Th}\) ions, leveraging established atomic physics techniques like laser cooling and quantum logic spectroscopy. While historically hindered by the lack of suitable lasers, The recent theoretical predication~\cite{xiao2024proposal, penyazkov2025theoretical} and the first experimental demonstration of continuous-wave VUV laser at 148.4 nm with sub-hertz linewidth and sufficient power~\cite{xiao2025cwVUV} now makes this approach a practical reality.

\begin{figure*}[htbp]
    \centering
    \includegraphics[width=0.85\textwidth]{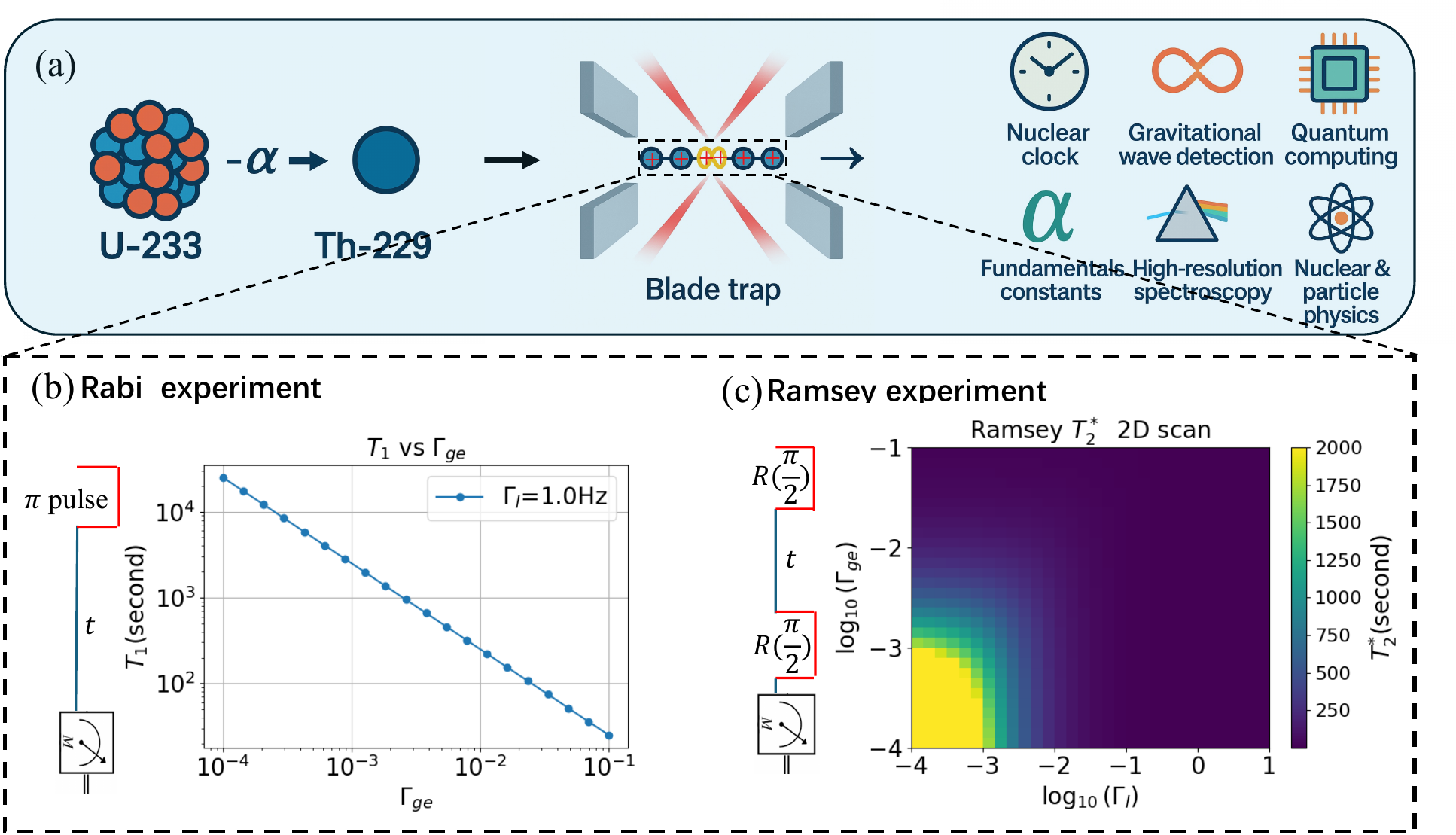} 
    \caption{We numerically simulated the dynamics of a $^{229}\mathrm{Th}$ nuclear-level qubit using currently accessible experimental parameters. (a) Schematic illustration of the experimental procedure: Th-229 is obtained from the decay of U-233, ionized, and subsequently loaded into a blade ion trap for entanglement operations. This platform offers a versatile framework with potential applications spanning nuclear clock development, gravitational-wave detection, and advanced quantum information processing. Specifically, we investigated the performance of both the $\pi$-pulse (Rabi) experiment and the Ramsey interference experiment under appropriate detuning and a driving laser power of 30~$\mu$W as show in (b) and (c). Numerical results indicate that, owing to the extremely low decay rate between the isomeric and ground nuclear states, the encoded qubit exhibits a very long $T_1$ time. Furthermore, if the phase noise of the VUV laser can be well controlled, the $T_2$ coherence time of the nuclear-spin-encoded qubit can also be greatly enhanced.} 
    \label{T1_T2} 
\end{figure*}

In this work, we theoretically analyze the potential of using $^{229}\mathrm{Th}^{3+}$ ions as direct nuclear-level qubits for quantum information processing in ion trap systems, focusing on scenarios enabled by current $148.4~\mathrm{nm}$ laser technology. We evaluate the performance of $^{229}\mathrm{Th}^{3+}$ ions as qubit carriers when trapped and manipulated via direct optical addressing. Furthermore, we investigate the feasibility of generating nuclear-state entanglement by employing red- and blue-detuned $148.4~\mathrm{nm}$ laser pulses, and quantify the achievable entanglement strength through pulse optimization. Finally, we discuss possible applications of nuclear-level entanglement in quantum metrology, nuclear clocks, and fundamental constant measurements. Our findings provide a feasible roadmap for developing quantum information protocols with nuclear isomers, highlighting the practical prospects of $^{229}\mathrm{Th}^{3+}$ ions under realistic experimental conditions.


\section{Theoretical Model and Qubit Encoding/Control in $\mathbf{^{229}Th^{3+}}$ Ions}

Trapped-ion quantum computing has witnessed remarkable experimental and theoretical progress in recent years, establishing itself as one of the leading platforms for universal quantum computation. In this section, we systematically describe the theoretical model underlying this study, including the level structure analysis of the triply ionized $^{229}\mathrm{Th}^{3+}$ ion and the proposed qubit encoding scheme.

\subsection{Electronic and nuclear structure of $\mathbf{^{229}Th^{3+}}$ and qubit encoding}


The $^{229}\mathrm{Th}$ atom has four valence electrons with a ground-state configuration of $6d^2 7s^2$. After removing three electrons, the resulting $^{229}\mathrm{Th}^{3+}$ ion retains a single valence electron, leading to a simplified electronic configuration of $5f^1$. This configuration results in a considerably simpler spectrum and exhibits high stability in the triply charged state, facilitating long-term manipulation in ion traps. Additionally, the relatively simple level structure enables straightforward laser cooling, notably using the 1088 nm transition (analogous to the D$_1$ line of rubidium (Rb) atoms)~\cite{yamaguchi2024laser}. 


Using conventional ion-trap techniques, triply charged $^{229}\mathrm{Th}^{3+}$ ions can be efficiently trapped, with experiments demonstrating stable confinement of 50--100 ions for over an hour~\cite{li2023scheme}.
The defining feature of $^{229}\mathrm{Th}$ is an unique low\mbox{-}energy nuclear isomeric transition at $\approx 8.4\,\mathrm{eV}$ ($\lambda \approx 148.4\,\mathrm{nm}$){~\cite{seiferle2019energy,tiedau2024laser,zhang2024frequency}; critically, continuous\mbox{-}wave VUV laser radiation at $148.4\,\mathrm{nm}$ has only recently become experimentally accessible~\cite{xiao2025cwVUV}.
We propose to encode the qubit in two hyperfine-resolved nuclear states within the same electronic 
manifold: the ground state $|5F_{5/2},I_g=5/2,F=5,m_F=5\rangle$ and the excited isomeric state $|5F_{5/2},I_e=3/2,F=4,m_F=4\rangle$. The isomeric excited state is characterized by a theoretically predicted radiative lifetime of $\sim 10^3$–$10^4$s~\citep{tkalya2015radiative,shigekawa2021estimation} , and its first-order sensitivities to magnetic fields, electric fields, and blackbody radiation are suppressed by 3–5 orders of magnitude compared to electronic states, naturally mitigating decoherence mechanisms~\cite{flambaum2006enhanced}.


\subsection{Qubit control and performance analysis}

Once nuclear isomeric states are chosen as the computational basis, a central challenge is to efficiently drive transitions between the ground and isomeric states. Such nuclear transitions, induced by electromagnetic interactions, are classified into electric or magnetic multipoles—e.g., electric quadrupole (E2) and magnetic dipole (M1)—since electric dipole (E1) transitions are forbidden~\cite{palffy2008electric}. M1 transitions are most pertinent, being highly sensitive to nuclear magnetic moments and spin configurations, and constrained by angular-momentum selection rules~\cite{bilous2018electric}. They enable precise probes of nuclear structure and form the foundation of nuclear clocks, which leverage the high stability of nuclear energy levels for timekeeping and fundamental tests~\cite{flambaum2006enhanced}. Transition rates follow from matrix elements expressed as spherical integrals over nuclear wavefunctions and electromagnetic fields~\cite{blatt2012theoretical}. For the $^{229}$Th isomer, the E2 probability is suppressed by $\sim 11$ orders of magnitude relative to M1~\citep{porsev2010effect}; we therefore consider only magnetic dipole-driven transitions when estimating control parameters.


Although the $^{229}$Th isomer has a relatively low excitation energy of 8.4~eV, it still exceeds typical electron- and nuclear-spin qubit energies. Two main approaches are proposed for driving this transition. The electron-bridge (EB) mechanism uses laser-driven electronic transitions to transfer energy and angular momentum to the nucleus, enabling efficient isomer population transfer~\cite{porsev2010effect,porsev2010electronic}. Alternatively, direct 148.4~nm VUV excitation—now feasible with continuous-wave sources offering sub-hertz linewidths~\cite{xiao2024proposal,penyazkov2025theoretical,xiao2025cwVUV}—provides a practical route for coherent nuclear-state control.

In this context, we propose employing direct laser excitation for quantum gate operations and the EB mechanism for qubit initialization and readout, leading to an effective Hamiltonian for the nuclear–field interaction.
\begin{align}
\hat{H'}=\frac{\Delta}{2}\sigma_{z}+\left(V_{ge}e^{-i\phi(t)}\sigma_x+V_{eg}e^{i\phi(t)}\sigma_x\right),
\label{eq:Ham}
\end{align}
where $\Delta = \omega_{l} - \omega_{a}$ represents the detuning between the laser frequency $\omega_l$ and the atomic transition frequency $\omega_a$, and $\phi(t)$ is the related phase of laser, $V_{eg}=V_{ge}^\ast=\int \mathbf{j}(\mathbf{r}) \cdot \mathbf{A}_0 e^{i \mathbf{k}_{\ell}\cdot \mathbf{r}} \, d^3 r$ is the driving field and $\mathbf{j}(\mathbf{r})$ is the nuclear current density, this driving field has relation both with the laser power and the energy decay rate which we can write as $V_{ge} =  \sqrt{\frac{I \pi \epsilon_0 \hbar \Gamma_{ge}}{k^3}}$~\citep{von2020theory}, where $\Gamma_{ge}$ is the excited state decay rate and $I$ is the laser intensity which has relation $I=\frac{P}{\pi w_0^2}$ with laser power $P$ and beam waist $w_0$, for which the detail  derivation refers to the Appendix~\ref{ap:Rabi}. Under such conditions, the atom is driven between the ground state and isomeric state at an effective Rabi frequency $\Omega_{ge} = 2V_{ge}$ and a setting detuning $\Delta$. While in order to calculate the qubit performance, we need to consider an open system and using Lindblad master equation 
\begin{align}
\dot{\rho} = -i[\hat{H}', \rho] + \mathcal{L}_{\text{decay}}[\rho] + \mathcal{L}_{\text{laser}}[\rho]
\label{eq:lindblad}
\end{align}
to calculate the qubit property, where the Lindblad term $\mathcal{L}_{\text{decay}}=\Gamma_{ge} \left( \hat{\sigma}_{ge} \rho \hat{\sigma}_{ge}^\dagger - \frac{1}{2} \{\hat{\sigma}_{ge}^\dagger \hat{\sigma}_{ge}, \rho\} \right)$ captures the dissipative processes and $\hat{\sigma}_{ge}$ is the jump operator and the laser phase noise Lindblad superoperator $\mathcal{L}_{\text{laser}}[\rho] = \frac{\Gamma_l}{2} 
\left( 2 \sigma_z \rho \sigma_z - 2\rho \right)$, the $\Gamma_l$ represent the control laser fluctuation, which has developed quickly in direct nuclear transition. For detailed formula forms, please refer to Appendix~\ref{ap:lindblad}.  

Recent advances in $^{229}\mathrm{Th}$ spectroscopy have enabled 148.4~nm VUV laser generation with proposed output powers above 30~$\mu$W~\citep{xiao2024proposal,penyazkov2025theoretical} and, for the first time, a continuous-wave source at this wavelength with sub-hertz linewidth~\cite{xiao2025cwVUV}, although the output power of this recent experiment was not yet optimized, the design outlined in Ref.~\cite{xiao2024proposal} indicates that sufficient laser power can be achieved for high-fidelity nuclear excitation. Using such state-of-the-art lasers and focusing optics (Table~\ref{tab:exp_parameters}), coherent driving of $^{229}\mathrm{Th}^{3+}$ nuclear transitions is now experimentally feasible.


Using the experimentally feasible driving parameters as shown in Table~\ref{tab:exp_parameters}, we apply Eq.~(\ref{eq:dot_rho_ge}) to numerically study the dependence of nuclear-qubit coherence—specifically the longitudinal relaxation time $T_1$ and transverse dephasing time $T_2$—on the laser phase noise amplitude $\Gamma_l$. We first determine the $\pi$-pulse duration for single-qubit gates by simulating coherent Rabi oscillations (Fig.~\ref{T1_T2}), confirming that a 148.4~nm VUV laser can complete a full Rabi cycle within milliseconds, enabling high-fidelity population inversion. We then evaluate $T_1$ and $T_2$ under varying control parameters using $\pi$-pulse relaxation and Ramsey interferometry. When the VUV laser is detuned by an appropriate offset, the $^{229}\mathrm{Th}^{3+}$ isomeric-state qubit exhibits long coherence times, underscoring the intrinsic resilience of its nuclear transitions to decoherence.




Building on the analysis in Fig.~\ref{T1_T2}, we further examine the $^{229}\mathrm{Th}^{3+}$ isomeric nuclear state's operational regime as a qubit by quantifying how coherence times vary with two key parameters: the intrinsic decay rate and the laser detuning $\Gamma_l$. Numerical scans of $T_1$ and $T_2$ across a broad parameter range reveal that $T_1$ is determined almost entirely by the natural isomeric decay, showing negligible sensitivity to $\Gamma_l$. In contrast, $T_2$ is strongly influenced by both intrinsic decay and laser-induced dephasing, indicating that long-lived coherence requires both intrinsic stability and high-quality external control. This behavior reflects a general principle: realizing a high-fidelity qubit demands favorable intrinsic properties coupled with precise, stable control to preserve quantum coherence.


\subsection{Rapid SPAM by electric bridge (EB) processing}\label{sec3}

When using nuclear energy levels as qubits, it is desirable for the energy broadening of the excited state to be minimal, as this ensures a long lifetime. However, during the processes of initial-state preparation and quantum-state measurement (SPAM), an excessively narrow energy broadening can lead to low rates of preparation and readout, thereby reducing the overall efficiency of the qubit. To address this issue, it is necessary to dynamically adjust the transition process from the excited state to the initial state. To achieve this, we propose using the EB technique, which employs intermediate electronic energy levels to allow rapid preparation and readout of $^{229}\mathrm{Th}^{3+}$ ions. The detailed derivation of the impact of the electronic bridge on the energy broadening of the excited state can be found in~\cite{porsev2010electronic} and the enhance rate of nuclear transition can be defined as $\gamma=\Gamma_{EB}/\Gamma_{ge}$, which can get the detail of derivation in Appendix~\ref{ap:EB}. 

\begin{figure}[t]
    \centering
    \includegraphics[width=0.5\textwidth]{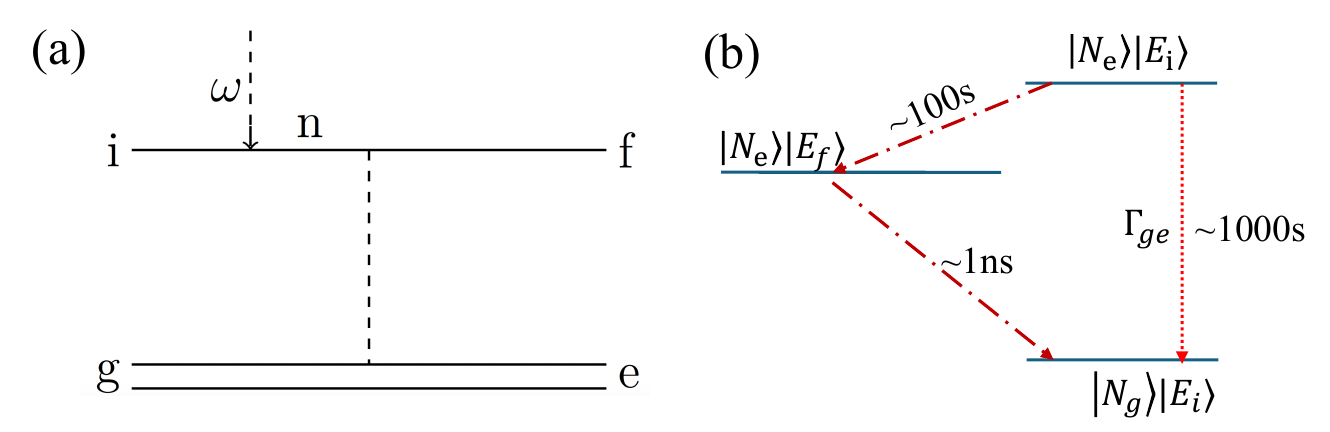} 
    \caption{Illustration of the electronic bridge (EB) process involved in nuclear-state transitions. (a) shows a schematic Feynman diagram of the EB process, adapted from~\cite{porsev2010electronic}, where the double solid line denotes the nuclear transition, the single solid line represents the electronic process, and the dashed line indicates the photon. (b) depicts the candidate states for our implementation together with their relevant parameters.} 
    \label{fig:eb} 
\end{figure}
 
The EB process can be illustrated by the Feynman diagram shown in Fig.~\ref{fig:eb}(a), the isomeric state \( |N_e\rangle|E_i\rangle \) is first driven to \( |N_e\rangle|E_f\rangle \) via an electronic excitation, without altering the nuclear state. The intermediate state \( |N_e\rangle|E_f\rangle \), having a short lifetime, then rapidly decays to the state \( |N_g\rangle|E_i\rangle \). In this mechanism, the overall decay rate of the nuclear isomer is predominantly governed by the laser-driven transition between the initial and intermediate electronic levels, which can be experimentally implemented at relatively high frequencies.

According to Ref.~\citep{dzuba2025resonance}, such an EB-assisted process can enhance the effective decay rate by 2 to 5 orders of magnitude. Furthermore, Ref.~\citep{seiferle2017lifetime} has experimentally demonstrated the possibility of preparing the nuclear ground state within 10~ms. These findings collectively indicate that the isomeric nuclear state is a viable candidate for encoding qubits, with feasible state initialization and readout on millisecond timescales.

\section{Entanglement of the nuclear isomeric states between $^{229}\mathrm{Th}^{3+}$ ions}

\subsection{Entanglement generation between nuclear isomeric states}

Even more intriguing than single qubit operation is the prospect of entangling the two nuclear states. In the mature field of trapped-ion quantum computing with electronic qubits, this is achieved by mediating interactions through the ions' collective motion (phonons). While the seminal Cirac-Zoller protocol first established this concept~\cite{cirac1995quantum}, it was the subsequent development of the Mølmer and Sørensen gate that provided a more robust and experimentally practical method~\cite{sorensen1999quantum}. 
The essential requirement for generating entanglement in Mølmer and Sørensen schemes is to realize an effective unitary of the form $U = e^{i\theta\sigma_\alpha^i \sigma_\alpha^j}$, where $\sigma_{\alpha}^{i}\!\in\!\{\sigma_{x},\sigma_{y},\sigma_{z}\}$ denotes the relevant spin operator determined by the laser configuration, and $\theta$ is the angle of entanglement, this unitary transformation constitutes the target operation for realizing nuclear entanglement in our system.
The central goal of this work is to demonstrate, for the first time, that this powerful, well-established framework can be translated from the electronic domain to the unique context of nuclear-level qubits.

\begin{table}[t]
\centering
\begin{tabular}{|l|c|}
\hline
\multicolumn{2}{|c|}{\textbf{Laser Parameters}} \\
\hline
Laser power                      & $P = 30~\mu\mathrm{W}$~\cite{xiao2024proposal}           \\
Laser wavelength                 & $148.3821~\mathrm{nm}$~\cite{tiedau2024laser}            \\
Red/Blue detuning                & $\Delta = 2.04~\mathrm{MHz}$      \\
Frequency drift tolerance        & $\delta\Delta = \pm 200~\mathrm{Hz}$~\cite{xiao2025cwVUV} \\
Laser beam waist                 & $w_0 = 1.5~\mu\mathrm{m}$           \\
Laser dutation time                 & $\tau = 10\text{--}100\mathrm{ms}$           \\
\hline
\multicolumn{2}{|c|}{\textbf{Trap Parameters}} \\
\hline
Phonon mode frequency                     & $\omega_{p} = [1.2, 2.08]\,\text{MHz}$ \\
Lamb-Dicke parameter matrix      & $\eta = \begin{pmatrix}
-0.1285 & -0.0976 \\
0.1285 & 0.0976
\end{pmatrix}$ \\
\hline
\end{tabular}
\caption{Key parameters for laser-driven entanglement of two $^{229}\mathrm{Th}^{3+}$ ions in a radio-frequency trap are summarized here. Under the limited power of the 148.4~nm laser~\cite{xiao2024proposal,tiedau2024laser}, the Rabi frequency is increased by tuning the detuning $\Delta$ close to a phonon mode (e.g., 2.08~MHz) and focusing the beam waist to $1$--$4~\mu\mathrm{m}$. This waist size enhances the local intensity while avoiding excessive sensitivity of the Rabi frequency to ion motion and minimizing crosstalk between neighboring ions spaced $5$--$10~\mu\mathrm{m}$ apart. Gate durations $\tau$ are set to $10$--$100~\mathrm{ms}$, matching current nuclear Rabi frequencies in the kHz range and allowing sufficient time to achieve entanglement within a single pulse. Phonon modes and Lamb--Dicke parameters are calculated using the standard second-quantization method described in Ref.\cite{leibfried2003quantum}.}
\label{tab:exp_parameters}
\end{table}


To accurately model phonon-mediated entanglement between nuclear-level qubits, it is essential to incorporate the small oscillatory motion of the nucleus within the ion trap into the laser–atom interaction formalism as shown in Eq.~(\ref{eq:Ham}). Then, the position-dependent terms in laser-atom interaction term can be written as
\begin{align}
\hat{V}_{ge}= \int \mathbf{j}(\mathbf{r}+\delta\mathbf{r}) \cdot \mathbf{A}_0 e^{i \mathbf{k}_{\ell}\cdot (\mathbf{r}+\delta\mathbf{r})} \, d^3 r 
= e^{ik\delta\mathbf{r}}V_{ge},
\end{align}
where $\delta\mathbf{r}$ represent the nuclear ion oscillation around the equilibrium position $\mathbf{r}$. This is due to the nuclear current density is related to the relative position of the nucleus, the relationship between $\mathbf{j}(\mathbf{r})$ and $\mathbf{r}$ remains the same formula~\cite{palffy2008electric}. The derivation here closely parallels that of conventional trapped-ion quantum gates. The main difference is that, with direct nuclear excitation enabled by the 148.4~nm laser, we can apply red- and blue-detuned sidebands to couple nuclear states via phonon modes—just as in standard electronic qubit systems. By determining the normal modes and Lamb–Dicke parameters for ${}^{229}\mathrm{Th}^{3+}$ ions, we identify the appropriate laser parameters for sideband transitions. This allows the small displacement term $k\delta \mathbf{r}$ to be quantized in the usual way, and the final multi-ion Hamiltonian takes the familiar form:
\begin{equation}
\hat{H}'' = \sum_{j,p} V_{ge}^j\, \eta_{j,p} \sin (\phi_j + (\Delta-\omega_p) t)\, \sigma_x^j (a_p + a_p^\dagger).
\label{eq:twoHam}
\end{equation}
Here, $\eta_{j,p}$ is the Lamb–Dicke parameter for the $j$-th ion and the $p$-th vibrational mode, $\omega_p$ is the frequency of the $p$-th collective phonon mode, and $a_p$, $a_p^\dagger$ are the annihilation and creation operators for that mode.

When the indices $j$ and $p$ in the Hamiltonian Eq.~(\ref{eq:twoHam}) each take multiple values, the internal states of the two nuclear levels are coupled via two phonon modes. Following the standard Mølmer–Sørensen framework for trapped-ion entangling gates~\cite{sorensen1999quantum}, a second-order Magnus expansion~\cite{blanes2009magnus} reveals that, by appropriately tuning laser power and relative phase to modulate $V_{ge}^j$ and $\phi_j$, then selecting a suitable detuning $\Delta$ and gate duration time $\tau$, the phonon modes can be completely decoupled from the ions within a single pulse cycle~\cite{manovitz2017fast}. Under these conditions, the resulting effective time-evolution operator—after eliminating the phononic degrees of freedom—is  
\begin{equation}
U = e^{i\pi\sigma_x^i \sigma_x^j/4},
\end{equation}
(see Appendix~\ref{ap:pulse} for derivation; the complete unitary including phonons is given in Appendix~\ref{ap:unitary}). When applied to the initial two-ion nuclear state $|00\rangle$, this operator yields $U|00\rangle = \frac{|00\rangle + i|11\rangle}{\sqrt{2}},$
demonstrating that tailored pulse modulation can generate maximally entangled nuclear states via phonon-mediated interactions.

While similar decoupling and gate-generation schemes have been extensively demonstrated for electronic-state qubits~\cite{benhelm2008towards,monroe2021programmable}, our work extends these methods to the optically addressable nuclear transition in $^{229}$Th. Compared to conventional trapped-ion platforms based on elements such as Yb$^+$, $^{229}\mathrm{Th}^{3+}$ ions in a trap exhibit relatively large inter-ion spacing, which for a two-ion system follows $d = \left(\frac{2e^{2}}{4\pi\epsilon_{0} m \omega_{z}^{2}}\right)^{1/3} Z^{2/3}$, where $Z$ is the charge number, without compromising the interaction strength. Furthermore, the 148.4~nm laser used to drive nuclear transitions offers a short wavelength and a correspondingly small Gaussian diffraction limit, with the beam waist scaling as $w_0\propto\lambda$ with laser wavelength $\lambda$, enabling higher-resolution individual ion addressing than longer-wavelength systems. This property also allows for the creation of tighter optical tweezers for phonon-spectrum engineering, which could benefit recently proposed schemes employing optical tweezers to scale up trapped-ion architectures~\cite{schwerdt2024scalable}. In addition, the large nuclear spin of $^{229}$Th provides a rich hyperfine manifold, enabling the use of multiple internal levels for qudit-based quantum information encoding. This approach combines the intrinsic coherence of nuclear states with the flexibility of phonon-mediated control, opening a new pathway for nuclear-clock-based quantum information processing.

To model decoherence, we include only the individual decay terms $\mathcal{L}_{\mathrm{decay}}$ for each ion and the laser-induced decoherence term $\mathcal{L}_{\mathrm{laser}}$ associated with the 148.4~nm excitation. The resulting Hamiltonian $\hat{H}''$ is then incorporated into the master equation in Eq.~(\ref{eq:lindblad}) to quantify the impact of these noise channels on the nuclear entanglement dynamics.


\begin{figure*}[t]
    \centering
    \includegraphics[width=0.95\textwidth]{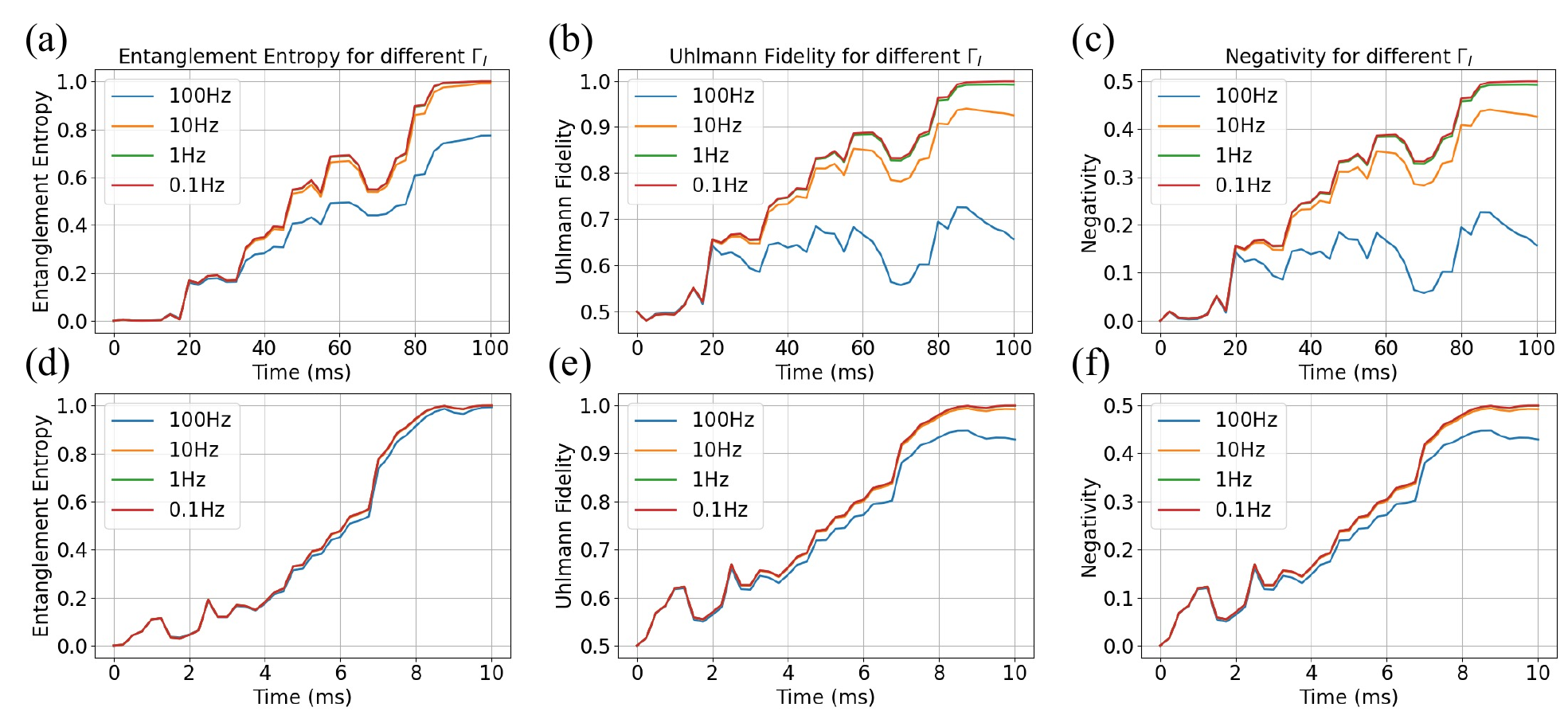} 
    \caption{Three complementary indicators of entanglement with different laser phase noise $\Gamma_l=[100, 10, 1, 0.1]\text{Hz}$, from left to right are entanglement entropy, Uhlmann fidelity and negativity, we employ a representative optimized pulse sequence, as illustrated in Fig.~\ref{fig:pulse}, to drive a two-ion system governed by the phonon-coupled Lindblad Eq.~(\ref{eq:lindblad}), and obtain the corresponding time-evolved density matrix. Entanglement entropy offers a direct measure of quantum correlations in a two-qubit pure state, with a value of 1 indicating maximal entanglement and 0 corresponding to a separable state. The Uhlmann fidelity reflects the overlap between the state evolved under the master equation and the target state; in this work, with the initial state $|00\rangle$ and the target Bell state, the fidelity starts at 0.5 and approaches 1 after the pulse sequence. Negativity quantifies entanglement for multi-qubit systems, and in the two-qubit case, a value near 0.5 signifies that the two qubits are maximally entangled. (Top) Simulation results with a maximum Rabi frequency of 20 kHz and gate duration time 100ms, corresponding to the baseline laser power. (Bottom) Results obtained with a $10\times$ increase in laser power, leading to a maximum Rabi frequency of 60 kHz. Increasing the laser power significantly shortens the required gate time to 10ms, thereby reducing the impact of qubit decoherence on nuclear-level entanglement.} 
    \label{fig:entanglement} 
\end{figure*}

\subsection{Numerical result of entanglement in nuclear isomeric states}\label{sec4}

We consider two $^{229}\mathrm{Th}^{3+}$ ions confined in a standard linear blade trap, with phonon modes and Lamb--Dicke parameters calculated via the second-quantization formalism of Ref.~\cite{leibfried2003quantum}. Using a second-order Magnus expansion of Eq.~(\ref{eq:twoHam}), we apply established optimization algorithms to determine pulse sequences as shown in Fig.~\ref{fig:pulse} that can realize maximize nuclear-state entanglement (see Appendix~\ref{ap:pulse}). Recent advances have demonstrated, for the first time, a continuous-wave 148.4~nm VUV source with sub-hertz linewidth~\cite{xiao2025cwVUV}. While this initial implementation was not optimized for output power, the theoretical design in Ref.~\cite{xiao2024proposal} suggests powers above 30~$\mu$W are realistically attainable. In our simulations, we adopt parameters from Ref.~\cite{xiao2024proposal} and assume a $1.5~\mu$m beam waist at the ion position, feasible with high-NA optics in a linear trap geometry. These parameters, listed in Table~\ref{tab:exp_parameters}, enable realistic estimates of coherent nuclear-level gate operations in $^{229}\mathrm{Th}^{3+}$ systems. However, the current spectral linewidth $\Gamma_l$ of the 148.4~nm laser remains relatively broad, so decoherence over the resulting hundreds-of-milliseconds gate durations $\tau$ is non-negligible and is explicitly included in our analysis and numerical simulations.


\begin{figure*}[htbp]
    \centering
    \includegraphics[width=0.95\textwidth]{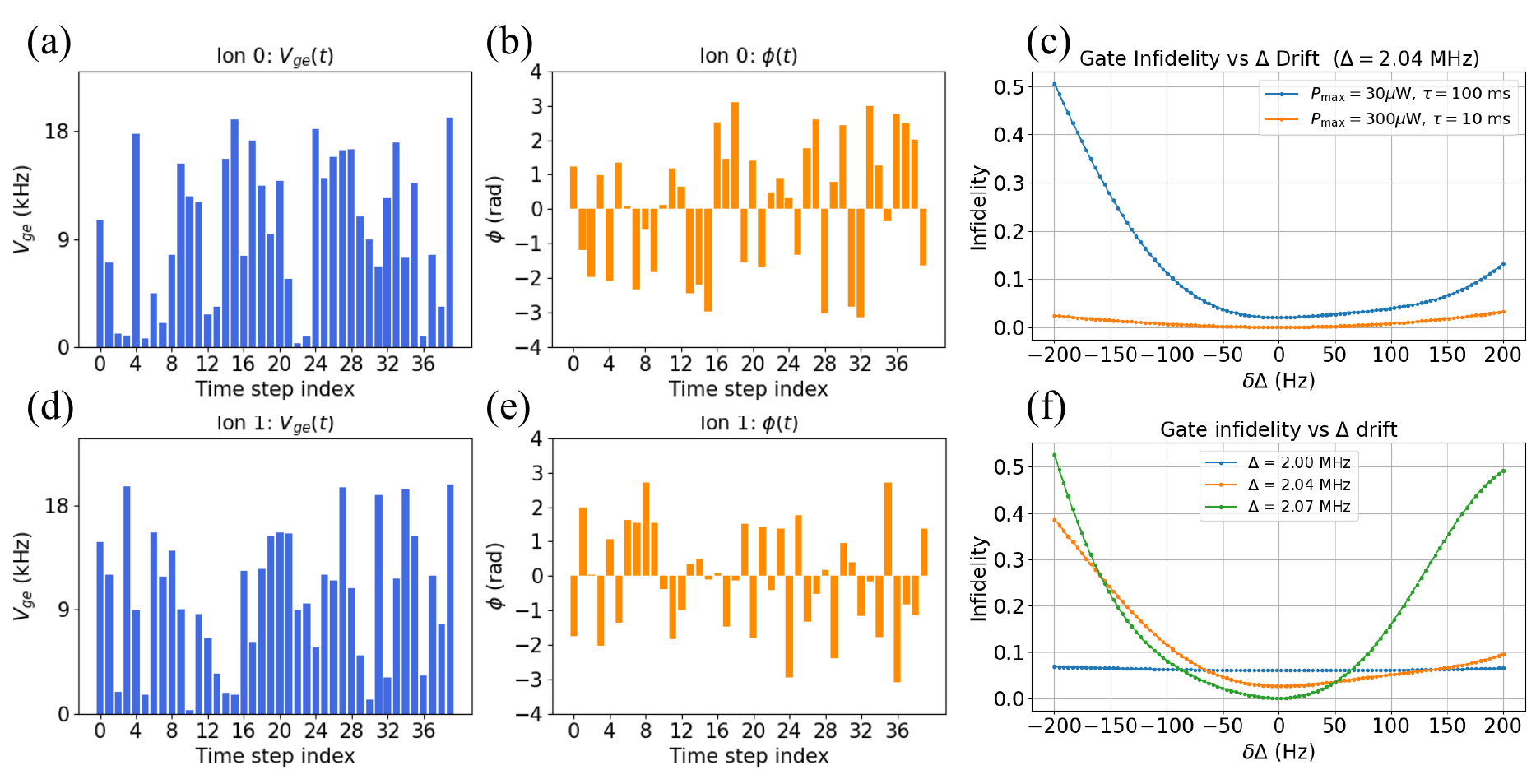} 
    \caption{Pulse sequences for two ions, here we choose amplitude(a,d) and phase(b,e) simultaneously modulation and set gate time to be $\tau=100$ms. In our numerical optimization, we use 40 pulse sequences and for every sequence, we restrict the effective Rabi frequency to between $3\sim 20$kHz corresponding to a maximum laser power of 30 $\mu$W. (c,f) Frequency robustness of the optimized entangling pulse under different laser powers $P_{max}=30\mu\text{W},300\mu\text{W}$ and detuning setting $\Delta=[2.0,2.04,2.07]\text{MHz}$. For a limited $148.4\,\mathrm{nm}$ laser power, the optimized pulse is highly sensitive to frequency drifts, with a tolerance window of only about $\delta\Delta=\pm 200\,\mathrm{Hz}$. In contrast, when the laser power is increased by an order of magnitude, the gate duration is significantly reduced, leading to much greater tolerance against frequency fluctuations. This demonstrates that enhanced laser power can effectively mitigate the adverse effects of technical noise on entanglement fidelity.}
    \label{fig:pulse} 
\end{figure*}

As a result, it is essential to account for decoherence by incorporating the Lindblad master equation for both ions. Specifically, we substitute the Hamiltonian $\hat{H}'$ in the Lindblad equation [Eq.~(\ref{eq:lindblad})] with the Hamiltonian $\hat{H}''$ from Eq.~(\ref{eq:twoHam}), and propagate the optimized control pulses through the Lindblad formalism to calculate the time evolution of the density matrix $\rho(t)$. Since our system consists of two ions each coupled to collective phonon modes, we obtain the reduced density matrix of the two-qubit system by tracing out the phonon degrees of freedom, i.e., $\rho_{AB} = \mathrm{Tr}_{\mathrm{phonon}}(\rho_{\mathrm{full}})$. To accurately characterize the dynamics of the entanglement between two nuclear states under these decoherence conditions, we evaluated three complementary indicators: \textit{entanglement entropy}, \textit{Uhlmann fidelity}, and \textit{negativity}~\cite{horodecki2009quantum,liang2019quantum} (see Fig.~\ref{fig:entanglement}). Each captures a different aspect of quantum correlations, enabling a thorough assessment of entanglement generation and loss under realistic conditions, including coupling to phonon modes. The three commonly used measures are the following:

\begin{align}
S &= -\mathrm{Tr}\left[\rho_A \log_2 \rho_A\right], \\
F &= \left[ \mathrm{Tr}\left( \sqrt{ \sqrt{\rho}\sigma\sqrt{\rho} } \right) \right]^2, \\
\mathcal{N} &= \frac{ ||\rho_{AB}^{T_B}||_1 - 1 }{2 }.
\end{align}
Here, $S$ is the entanglement entropy computed from the reduced density matrix $\rho_A = \mathrm{Tr}_B(\rho{_{AB}})$~\cite{wootters1998entanglement}; $F$ is the Uhlmann fidelity~\cite{uhlmann1976transition}, directly measuring the overlap between the simulated two-qubit state $\rho$ and the target state $\sigma$; and $\mathcal{N}$ is the negativity~\cite{vidal2002computable}, which quantifies bipartite entanglement based on the Peres-Horodecki criterion~\cite{peres1996separability, horodecki1996necessary}. While entanglement entropy effectively characterizes quantum correlations between the ions for pure states, it is relatively insensitive to entanglement leakage into the phonon modes. In contrast, both Uhlmann fidelity and negativity are more sensitive to such leakage and provide a more accurate assessment of genuine two-qubit entanglement in realistic settings. By utilizing all three measures, we can monitor the creation and loss of entanglement, distinguishing genuine quantum correlations from those masked by residual coupling to phonon modes—an essential approach for evaluating quantum gate performance in experiments.

As shown in Fig.~\ref{fig:entanglement}(a--c), we initially prepare the system in the $|00\rangle$ state, with the goal of generating the maximally entangled Bell state $|\psi_{\mathrm{Bell}}\rangle = \frac{1}{\sqrt{2}}(|00\rangle + i|11\rangle)$. By applying the optimized pulse sequences to Eq.~(\ref{eq:twoHam}) and limiting the 148.4~nm laser power to below 30~$\mu$W, we observe that, under various laser linewidths $\Gamma_l$, the entanglement entropy, Uhlmann fidelity, and negativity all increase at first as the pulse is applied. However, as the gate duration becomes longer, decoherence effects gradually dominate the system dynamics, leading to a sharp decrease in Uhlmann fidelity and negativity, as shown in Fig.~\ref{fig:entanglement}(b,c). This degradation is more pronounced for larger $\Gamma_l$, indicating that decoherence cannot be neglected in the current regime of nuclear-level entanglement, where limited laser power necessitates long gate times. Looking ahead, if future technological advances enable significantly higher 148.4~nm laser power---e.g., a 10$\times$ increase---the required gate time could be shortened by an order of magnitude. We simulate this scenario in Fig.~\ref{fig:entanglement}(d--f), showing the dynamical evolution of all three entanglement measures under the same conditions. In this regime, decoherence effects are significantly reduced, resulting in markedly improved entanglement performance at the nuclear level.

To further understand the effect of laser power on experimental feasibility, we analyze the frequency robustness of the optimized entangling pulses under different laser power constraints as show in Fig~\ref{fig:pulse}. A typical optimized pulse for two nuclear ions as show in Fig~\ref{fig:pulse}, When the available $148.4\,\mathrm{nm}$ laser power is limited, the laser frequency must be tuned very close to a specific motional sideband to achieve strong coupling, such as a detuning of $2.04\,\mathrm{MHz}$ near the $2.08\,\mathrm{MHz}$ phonon mode in our simulations. This configuration renders the entangling process highly sensitive to frequency drifts, restricting the robustness window to approximately $\pm200\,\mathrm{Hz}$. This sensitivity is attributable to the long interaction time imposed by low laser power, which amplifies the impact of technical noise. As shown in Fig.~\ref{fig:pulse}(a), when the laser power is increased by an order of magnitude, the required gate time is greatly reduced and the entanglement protocol becomes much more tolerant to frequency fluctuations. These results highlight that increasing the available laser power not only mitigates decoherence but also substantially enhances the experimental robustness of nuclear-level entanglement protocols.

\section{Discussion and conclusion}\label{sec13}

The nuclear isomeric state of $^{229}\mathrm{Th}$---the lowest-energy optically addressable nuclear transition known---combines high nuclear-level stability with atomic-level optical accessibility~\cite{tiedau2024laser}, while in its triply charged form it also offers practical advantages for scalable quantum computing: larger inter-ion spacing reduces optical crosstalk, the short-wavelength (148.4~nm) control laser facilitates high-resolution individual addressing, and the large nuclear spin provides a rich hyperfine manifold for multi-level qudit encoding. These features make $^{229}\mathrm{Th}$ nuclear-level qubits a promising basis for \textit{potentially} meaningful advances in both quantum information processing and ultra-high-precision metrology~\cite{campbell2011wigner,ludlow2015optical,von2020229,thirolf2024thorium}.

In precision measurement, trapping $^{229}\mathrm{Th}^{3+}$ ions in a linear blade ion trap and generating high-fidelity multi-qubit entanglement could \textit{potentially} enable Heisenberg-limited phase sensitivity~\cite{huelga1997improvement,leibfried2004toward,kessler2014heisenberg,huang2024entanglement}, with the \textit{potential} to improve gravitational-wave detection by isolating low-frequency signals from electromagnetic noise via the intrinsic shielding of nuclear transitions~\cite{campbell2012single,kolkowitz2016gravitational}. Likewise, a quantum nuclear clock built on entangled $^{229}\mathrm{Th}$ ion chains would feature frequency uncertainty scaling as $1/N$, \textit{potentially} achieving stability at the theoretically enabling $10^{-20}$ level and setting new benchmarks for global navigation and deep-space communication~\cite{hemmati2006deep,giorgi2019advanced}.

For probing fundamental physics, the $^{229}\mathrm{Th}$ transition offers notable sensitivity to variations in the fine-structure constant $\alpha$---with sensitivity factors exceeding $10^{4}$, three to four orders beyond atomic transitions~\cite{quercellini2012real,peik2021nuclear,derevianko2022fundamental,beeks2024fine}. Entanglement-enhanced dual-channel nuclear spectroscopy could \textit{potentially} boost detection of $\alpha$ drifts and proton charge radius shifts by over an order of magnitude, enabling stringent tests of time-varying constants and dark-energy coupling models~\cite{derevianko2014hunting,stadnik2015searching}. Multi-body entanglement would further allow precise determination of nuclear multipole moments, illuminating shell structure and collective deformation phenomena in heavy nuclei, while squeezed-noise nuclear spectroscopy could reduce frequency uncertainties to the $\mu$Hz level, \textit{potentially} resolving subtle molecular perturbations and cosmological redshift drifts~\cite{kotler2014measurement,safronova2018search,colombo2022entanglement,arrowsmith2024opportunities,martins2024cosmology}.

In quantum information processing, combining long-coherence nuclear-level qubits with optically addressable electronic-shell qubits enables dual-timescale hybrid encoding—supporting both extended quantum memory and rapid quantum logic operations, thereby expanding the design space for quantum error correction, scalable architectures, and distributed quantum networks~\cite{nakamura2024hybrid,chang2025hybrid}. Leveraging their high intrinsic coherence, long-lived nuclear-state qubits are ideal hosts for logical qubits in quantum error-correcting codes; by generating multi-qubit entanglement across such qubits, QEC can actively correct residual decoherence and operational errors~\cite{shor1995scheme}. In precision metrology, this protection preserves the narrow nuclear transition linewidth while maintaining signal sensitivity, enabling frequency estimation at or near the Heisenberg limit even under realistic noise~\cite{kessler2014quantum,zhou2018achieving,zhou2021asymptotic,wang2022quantum,faist2023time}.

These applications are now within reach due to enabling advances, particularly the first continuous-wave VUV laser at 148.4~nm with sub-hertz linewidth and adequate power, which permits direct optical control and entanglement of nuclear states. The ultra-narrow linewidth and long coherence of $^{229}\mathrm{Th}^{3+}$ transitions, together with frequency-comb and ion-trap technologies, enable high-fidelity qubit manipulation and efficient phonon-mediated entanglement. Theoretical modeling indicates scalability, positioning $^{229}\mathrm{Th}^{3+}$ nuclear-level qubits as a \textit{potentially} foundational platform for next-generation quantum processors, high-precision clocks, and fundamental-physics experiments.


\begin{acknowledgements}
This work is supported by the National Natural Science Foundation of China (Grant No.~92365111), Shanghai Municipal Science and Technology (Grant No.~25LZ2600200), Beijing Natural Science Foundation (Grants No.~Z220002), and the Innovation Program for Quantum Science and Technology (Grant No.~2021ZD0302400).
\end{acknowledgements}

\appendix
\section{The Rabi frequency relation with the nuclear excited state decay rate}\label{ap:Rabi}
In the derivation above, the Rabi frequency can be defined in the form $\Omega_{ge} = 2\langle e|V_{eg}\sigma_x|g\rangle=2|V_{ge}|$. For magnetic dipole transitions in nuclear energy levels, the integral of $r$ can be explicitly evaluated through spherical surface integration, with its specific expression given as follows~\cite{palffy2008electric}:

\begin{align}
\Omega_{ge}&= B_0 \sqrt{2\pi} \sqrt{\frac{L+1}{L}} \frac{k^{L-1}}{(2L+1)!!} 
\left| \langle e | \hat{\mathcal{M}}_{L \sigma} | g \rangle \right|,
\end{align}
where $L$ and $\sigma$ denote the state angular momentum and magnetic quantum number, and $\hat{\mathcal{M}}_{L \sigma}$ is the magnetic multipole operator, which can be written as~\cite{ring2004nuclear}:

\begin{align}
\hat{\mathcal{M}}_{L \sigma} = \frac{1}{L+1} \int d^3 r \, \left[ \mathbf{r} \times \mathbf{j} \right] \nabla \left( r^L Y_{L \sigma} \right),
\end{align}
where $Y_{L \sigma}$ are the spherical harmonics and $\mathbf{j(r)}$ is the current density in nucleus atom.

It should be noted that the radiative decay rate from the excited state $|e\rangle$ to the ground state $|g\rangle$ can be expressed using the following formula:

\begin{align}
\Gamma_{ge} = \frac{2 \mu_0}{\hbar} \frac{L+1}{L \left[ (2L+1)!! \right]^2} k^{2L+1} \left| \langle g | \hat{\mathcal{M}}_{L \sigma} | e \rangle \right|^2,
\end{align}
so the Rabi frequency has relation with radiative decay rate as:

\begin{align}
\Omega_{ge} 
= 2V_{ge} = 2\langle e|H_I | g\rangle =  \sqrt{\frac{E_0^2 \pi \epsilon_0 \hbar \Gamma_{ge}}{k^3}}.
\end{align}

\section{The lindblad equation in nuclear transition}\label{ap:lindblad}

Because the $^{229}\mathrm{Th}^{3+}$ is insensitive with the environment noise, for simply, we just consider the isomeric state decay $\Gamma_{ge}$ and control laser fluctuation, than the density matrix of the two level system can be written as~\citep{von2020theory}:

\begin{align}
\dot{\rho}_{ee} &= -\dot{\rho}_{gg} = -i \left[ V_{eg} \rho_{ge} - V_{ge} \rho_{eg} \right] - \rho_{ee} \Gamma_{ge}, \notag\\
\dot{\rho}_{ge} &= \dot{\rho}_{eg}^* = -i V_{ge} \left[ \rho_{ee} - \rho_{gg} \right] - (i\Delta + \tilde{\Gamma})\rho_{ge},
\label{eq:dot_rho_ge}
\end{align}
where $\tilde{\Gamma}=\frac{(\Gamma_l+\Gamma_{ge})}{2}$. In condition where $\Omega_{ge}\gg \max(\Gamma_{ge}/2,\Gamma_l/2)$ and the system starts in the ground state at $t = 0$, the population of the excited state, $\rho_{ee}(t)$, under resonance conditions can be can be obtained analytically:
\begin{align}
\rho_{ee}(t) =& \frac{\tilde{\Gamma}\Omega_{eg}^2}{2 \left( \Gamma_{ge}(\Delta^2 + \tilde{\Gamma}^2)  + \tilde{\Gamma}\Omega_{eg}^2 \right)}\\ \notag 
&\times \left[ 1 - e^{-\frac{1}{2} (\Gamma_{ge} + \tilde{\Gamma}) t} 
\left( \cos(\lambda t) + \frac{\Gamma_{ge} + \tilde{\Gamma}}{2\lambda} \sin(\lambda t) \right) \right],
\end{align}
the solution shows Rabi oscillations damped by the decay rates, where $\lambda=|\Omega_{ge}^2 + \Delta^2 -(\tilde{\Gamma} + \Gamma_{ge})^2/4|^{1/2}$ is the modified oscillation frequency. The excited state population oscillates with frequency $\Omega_{ge}$ and decays exponentially due to the Lindblad dissipation.

\section{Electric bright process in nuclear transition}\label{ap:EB}
The nuclear transition probability enhanced by the EB process can be obtained from the following derivation:
\begin{align}
\Gamma_{\mathrm{EB}} \approx& \left(\frac{\omega}{c}\right)^3
\frac{\left|\langle I_g \| \mathcal{M}_1 \| I_e \rangle\right|^2}{(2I_e+1)(2J_i+1)} \notag \\
&\sum_{J_n}\frac{1}{[J_n]}\left|\sum_{\gamma_n}\frac{\langle \gamma_i J_i\|\mathcal{D}_1\|\gamma_n J_n\rangle\langle \gamma_n J_n\|\mathbf{H_{int}}\|\gamma_f J_f\rangle}{\omega_{in}-\omega_N}\right|^2,
\label{eq:eb}
\end{align}
the parameter $\omega$ is the photon frequency which we be written $\omega=\epsilon_i-\epsilon_f+\omega_N$, and $\mathcal{M}_1$ denotes the nuclear M1 transition operator. The quantum numbers $I_g$, $I_m$, and $J_i$ represent the nuclear spins of the ground nuclear state, the isomeric nuclear state, and the total electronic angular momentum of the initial electronic state and $[J_n] = 2J_n + 1$, respectively. The quantities $\gamma_i J_i$, $\gamma_n J_n$, and $\gamma_f J_f$ represent the initial, intermediate, and final electronic states, respectively, specified by their electronic quantum numbers ($\gamma$) and the total electronic angular momentum ($J$). The operator $\mathcal{D}_1$ is the electric dipole moment, whereas $\mathbf{H}_{\mathrm{int}}$ symbolizes the hyperfine interaction of Hamiltonian coupling electronic and nuclear degrees of freedom. The transition frequencies $\omega_{in}$ and $\omega_N$ reflect, respectively, the energy difference between electronic states, the nuclear transition frequency, highlighting the resonance conditions critical for the EB process.

\section{Cost function to optimize the pulse for nuclear state entanglement}\label{ap:pulse}

We determine the frequencies $\omega_p$ of the collective vibrational modes—phonons—within the ion chain, as well as the corresponding displacements from equilibrium $b_{j,p}$. These frequencies and displacements allow us to quantify the small vibrations around the equilibrium positions. Thus, 
$k\delta r_{j}$ can be second quantized as:
\begin{align}
k\delta r_{j}\rightarrow\sum_{p} kb_{j,p}(a_p e^{i\omega_{p}t} + a_p^{\dagger} e^{-i\omega_{p}t}),
\end{align}

where $\eta_{j,p}=k b_{j,p}$ is referred to as the Lamb-Dicke parameter, and $a_p,a_p^\dagger$ are the annihilation and creation operators of phonons. Using these definitions, we rewrite equation~(1) in the main text with second quantization as:
\begin{align}
\hat{H}'=\frac{\Delta}{2}\sigma_{z} +& V_{ge}e^{-i\big(\phi_{j}(t)+\sum_{p}\eta_{j,p}(a e^{i\omega_{p}t}+a^{\dagger}e^{-i\omega_{p}t})\big)}\sigma_{x} \notag \\
+& V_{eg}e^{i\left(\phi_{j}(t)+\sum_{p} \eta_{j,p}(a _pe^{-i\omega_{p}t}+a_p^{\dagger}e^{i\omega_{p}t})\right)}\sigma_{x}.
\end{align}
In multiple ions with blue/red side laser, the Hamiltonian can be simplify to:
\begin{align}
\hat H'' = \sum_{j,p}V_{ge}^j\eta_{j,p}\sin (\phi_j+\delta_p t)\sigma_x^j (a_p+a_p^\dagger),
\label{phonon_ion}
\end{align}
where $\delta_p=\Delta-\omega_p$ and we can evolve with time $\tau$ and expand to the second order of Magnus in Lamb-Dicke regime.
\begin{align}
U(\tau)=e^{i\big(\sum_{j,p}\beta_{j,p}(\tau)\sigma_x(a_p+a_p^\dagger)+\sum_{j,j'}\theta_{j,j'}\sigma_x^j\sigma_x^{j'}\big)},
\label{ap:unitary}
\end{align}
The first term represents the coupling between the internal states of the ions and the phonon modes which is:
\begin{align}
\beta_{j,p}(\tau) =\int_{0}^{\tau} dt\, \eta_{j,p}\,V_{ge}^j(t) \sin(\phi_j(t)+\delta_p t),
\end{align}
while the second term corresponds to an effective coupling between internal states of different ions which is:
\begin{align}
\theta_{j,j'}(\tau) =&\sum_p \int_{0}^{\tau}\int_{0}^{t_{1}} dt_{1} dt_{2}\,\eta_{j,p}\eta_{j',p}V_{ge}^j(t_{1})V_{ge}^{j'}(t_{2})\nonumber\\
&\sin(\phi_j(t_1)+\delta_p t_1)\sin(\phi_{j'}(t_2)+\delta_p t_2).
\end{align}
When the first term $\beta_{j,p}(\tau)$ vanishes, the phonons effectively decouple from the internal states, and if $\theta_{j,j'}(\tau)\neq0$, the internal states of different ions remain coupled, enabling entanglement generation. By choosing appropriate time-dependent control parameters $(V_{ge}^j(t), \phi_{j}(t))$, one can realize conditions under which phonon coupling is eliminated while ion-ion entanglement is maximized, so we define the cost function as:

\begin{equation}
\mathcal{C} = \sum_{j,p} \left( |\beta_{j,p}(\tau)|^2 
+ \left| \sum_{j \neq j'} \theta_{j,j'}(\tau) - \frac{\pi}{4} \right| \right),
\label{eq:cost}
\end{equation}
and determine the pulse parameters required to achieve an MS gate with 99.9\% fidelity. \\


\bibliographystyle{apsrev4-1}
\bibliography{references}

\end{document}